\documentclass[prl,letterpaper,twocolumn,tightenlines,superscriptaddress,showpacs,preprintnumbers,nofootinbib,floatfix]{revtex4}

\usepackage{amsmath,amssymb,amsfonts,graphicx,pifont,dcolumn}

\RequirePackage{slashed}

\def\a{\alpha}

\def\s{\sigma}
\def\D{\Delta}

\def\lra{\longrightarrow}
\def\EA{{\textit{et al.}}}

\def\lf{\left}
\def\rg{\right}
\def\la{\langle}
\def\ra{\rangle}

\begin{document}

\preprint{JLAB-THY-09-1051,~NT@UW-09-19}

\author{W.~Bentz}
\affiliation{Department of Physics, School of Science, Tokai University,
             Hiratsuka-shi, Kanagawa 259-1292, Japan}

\author{I.~C.~Clo\"et}
\affiliation{Department of Physics, University of Washington, Seattle, WA 98195-1560, USA}

\author{J.~T.~Londergan}
\affiliation{Department of Physics and Nuclear Theory Center
             Indiana University, Bloomington, IN 47405, USA}

\author{A.~W.~Thomas}
\affiliation{CSSM, School of Chemistry and Physics, University of Adelaide, 
Adelaide SA 5005, Australia, \\ 
and Jefferson Lab, 12000 Jefferson Avenue, Newport News, VA 23606 USA 
}

\title{Reassessment of the NuTeV determination of the weak mixing angle}

\begin{abstract}
In light of the recent discovery of the importance of the isovector EMC
effect for the interpretation of the NuTeV determination of $\sin^2 \theta_W$,
it seems timely to reassess the central value and the errors on this 
fundamental Standard Model parameter derived from the NuTeV data.
We also include earlier work on charge symmetry violation and 
the recent limits on a possible asymmetry between $s$ and $\bar{s}$ quarks.
With these corrections we find a revised NuTeV result of
$\sin^2 \theta_W = 0.2221 \pm 0.0013(\text{stat}) \pm 0.0020(\text{syst})$,
which is in excellent agreement with the running of $\sin^2 \theta_W$ predicted 
by the Standard Model. As a further check, we find that the separate ratios of neutral 
current to charge current cross sections for neutrinos and for antineutrinos are both 
in agreement with the Standard Model, at just over one standard deviation, once 
the corrections described here are applied.
\end{abstract}

\maketitle

Using a very careful comparison of the charged and neutral current total cross 
sections for $\nu$ and $\bar{\nu}$ on an iron target, 
the NuTeV collaboration reported a three standard deviation discrepancy 
with the Standard Model value of $\sin^2 \theta_W$~\cite{Zeller:2001hh}. 
This was initially taken as an indication of possible new physics,
however attempts to understand this anomaly in terms of popular extensions of 
the Standard Model have proven unsuccessful~\cite{Davidson:2001ji,Kurylov:2003by}.
At the same time a number of possible corrections within the Standard Model have been 
suggested~\cite{Londergan:2003ij,Martin:2003sk,Martin:2004dh,Gluck:2005xh,Mason:2007zz,Ball:2009mk,Cloet:2009qs}, 
most of which have a sign likely to reduce this discrepancy. 

The correction associated with charge symmetry violation (CSV), 
arising from the $u$- and $d$-quark mass differences~\cite{Sather:1991je,Rodionov:1994cg},
has been shown to be largely model independent 
and to reduce the discrepancy by about 1$\sigma$~\cite{Londergan:2003ij}. 
If the momentum fraction 
carried by $s$-quarks in the proton exceeds that carried by $\bar{s}$-quarks, as 
suggested by chiral physics~\cite{Signal:1987gz,Thomas:2000ny} and recent experimental 
analysis~\cite{Mason:2007zz}, there could be a 
further reduction, albeit with large uncertainties at present~\cite{Ball:2009mk}. 
Finally, in Ref.~\cite{Cloet:2009qs} it was recently pointed out 
that the excess neutrons in iron lead to 
an isovector EMC effect that modifies the parton distribution functions (PDFs) 
of \textit{all} the nucleons in the nucleus.
Qualitatively this has the same sign as the CSV correction and a quantitative 
estimate suggests that it reduces the NuTeV discrepancy with the Standard Model by 
about 1.5$\sigma$~\cite{Cloet:2009qs}. 

These effects are essentially independent and can therefore be combined in a 
straightforward manner. It is then immediately clear that the corrected NuTeV 
data will be more consistent with the Standard Model.
Rather than continuing to report that the data is $3 \sigma$ above expectations, 
we suggest that it is timely to update the derived value of $\sin^2 \theta_W$.

In this Letter we will examine the corrections in turn, 
assign to each a central value
and a conservative error, then combine them to produce a revised value for
$\sin^2 \theta_W$. Our final value is
\begin{align}
\sin^2 \theta_W = 0.2221 \pm 0.0013(\text{stat}) \pm 0.0020(\text{syst}),
\end{align}
which is in excellent agreement with the corresponding 
Standard Model result, namely $0.2227 \pm 0.0004$~\cite{Abbaneo:2001ix,Zeller:2001hh} in the on-shell 
renormalization scheme. The original NuTeV result was
$\sin^2\theta_W = 0.2277 \pm 0.0013(\text{stat.}) \pm 0.0009(\text{syst.})$~\cite{Zeller:2001hh}.

The NuTeV experiment involved a precise measurement on a steel target of the ratios 
$R^\nu$ and $R^{\bar{\nu}}$, which are the ratios of the neutral current (NC) to charged 
current (CC) total cross sections for $\nu$ and $\bar{\nu}$, respectively.
Integral to the NuTeV extraction of $\sin^2 \theta_W$ was a detailed Monte Carlo 
simulation of the experiment. However, NuTeV have provided functionals which allow one to accurately estimate the effect of any proposed correction \cite{Zeller:2002du}.

The NuTeV study was motivated by the observation of Paschos and Wolfenstein 
\cite{Paschos:1972kj} that a ratio of cross sections for neutrinos 
and antineutrinos on an isoscalar target allowed an independent extraction 
of the weak mixing angle. The so-called Paschos-Wolfenstein (PW) ratio 
is given by~\cite{Paschos:1972kj,paschos}
\begin{equation}
R_{\text{PW}} = \frac{\s_{NC}^{\nu\,A} - 
\s_{NC}^{\bar{\nu}\,A}}{\s_{CC}^{\nu\,A} - \s_{CC}^{\bar{\nu}\,A}} 
\equiv \frac{R^{\nu} - r R^{\bar{\nu}}}{1 - r} \ .   
\label{eq:PW}
\end{equation}
In Eq.~(\ref{eq:PW}), $R_{\text{PW}}$ is the PW ratio, $A$ represents the 
nuclear target, and $r=\s_{CC}^{\bar{\nu}\,A} / \s_{CC}^{\nu\,A}$.
Expressing the 
total cross-sections in terms of quark distributions, ignoring the heavy 
quark flavours and $\mathcal{O}(\a_s)$ corrections, the PW ratio becomes
\begin{align}
R_{\text{PW}} = \tfrac{\lf(\tfrac{1}{6}-
\tfrac{4}{9} s^2_W\rg) \lf\la x_A\,u^-_A\rg\ra 
        + \lf(\tfrac{1}{6}-\tfrac{2}{9} s^2_W\rg)\, \lf\la x_A\,d^-_A + x_A\,s^-_A\rg\ra}
       {\lf\la x_A\,d^-_A  + x_A\,s^-_A\rg\ra - \tfrac{1}{3}\lf\la x_A\,u^-_A \rg\ra},
\label{eq:PW_quark}
\end{align}
where $s^2_W \equiv \sin^2\theta_W$, $x_A$ is the Bjorken scaling variable for the 
nucleus multiplied by $A$, $\la \ldots \ra$ implies integration over $x_A$,
and $q_A^- \equiv q_A - \bar{q}_A$ are the non-singlet quark 
distributions of the target.
 
Ignoring quark mass differences, strange quark effects and 
electroweak corrections, the $u$- and $d$-quark distributions of an isoscalar 
target will be identical, and in this limit Eq.~\eqref{eq:PW_quark} becomes
$R_{\text{PW}} \stackrel{N=Z}{\lra} \tfrac{1}{2} - \sin^2\theta_W$.
If corrections to this result are small the PW ratio provides 
an independent determination of the weak mixing angle. 
Expanding Eq.~\eqref{eq:PW_quark} about the $u^-_A = d^-_A$ and 
$s_A^- \ll u^-_A + d^-_A$ limits, we obtain the leading PW correction term, namely
\begin{align}
\D R_{\text{PW}} \simeq \lf(1-\frac{7}{3}s^2_W\rg)
                \frac{\la x_A\,u^-_A - x_A\,d^-_A - x_A\,s^-_A\ra}{\la x_A\,u^-_A + x_A\,d^-_A\ra}.
\label{eq:correction}
\end{align}
Extensive studies of neutrino-nucleus reactions have concluded that the most 
important contributions to Eq.~\eqref{eq:correction} arise from nuclear 
effects, CSV and strange quarks. These corrections will be discussed in turn 
below. 

In discussing the extraction of the weak mixing angle from neutrino reactions, 
it is customary and pedagogically useful to refer to corrections to the PW 
ratio, and we follow this practice. However, it is important to remember that 
in the NuTeV analysis the measured quantities were the NC to CC ratios for 
neutrinos and antineutrinos, and that the weak mixing angle was extracted 
through a Monte Carlo analysis. For a given effect, the PW ratio will give 
only a qualitative estimate of the correction to the weak mixing angle. 
Quantitative corrections are obtained by using the functionals provided 
by NuTeV \cite{Zeller:2002du}. Throughout this work we denote a contribution 
to Eq.~\eqref{eq:correction} by $\D R^i_{\text{PW}}$, while 
the best estimate of the correction to the NuTeV determination of $\sin^2\theta_W$, 
calculated using a NuTeV functional, 
is denoted by $\D R^i \equiv \D^i \sin^2\theta_W$, where, in each case, $i$ 
labels the type of correction. For completeness, at the end of our discussion 
we also report the effect of the corrections which we have considered on the 
separate ratios $R^\nu$ and $R^{\bar{\nu}}$.

%===============================================================================
%###############################################################################
%===============================================================================
\textit{Nuclear corrections} ---
For sufficiently large $Q^2$, nuclear corrections to the PW ratio for an 
isoscalar nucleus are thought to be negligible. 
However, the NuTeV experiment was performed on a steel target 
and it is essential to correct for the neutron excess before extracting $\sin^2\theta_W$.
NuTeV removed the contribution of the excess neutrons to the cross-section
by assuming that the target was composed of free nucleons. 

However, the recent results of 
Clo\"et \textit{et al.}~\cite{Cloet:2009qs} have shown 
that the excess neutrons in 
iron have an effect on \textit{all} the nucleons 
in the nucleus, which is not accounted for by a subtraction of 
their naive contribution.
In particular, the isovector-vector mean-field generated by the difference in proton 
and neutron densities, $\rho_p(r) - \rho_n(r)$, acts on every $u$- and $d$-quark 
in the nucleus and results in the break down of the usual assumption that
$u_p(x) = d_n(x)$ and  $d_p(x) = u_n(x)$ for the bound nucleons.
An explicit calculation of this correction was made in Ref.~\cite{Cloet:2009qs}, using 
the approach of Bentz \textit{et al.}~\cite{Mineo:2003vc,Cloet:2005rt,Cloet:2006bq}.

The correction associated with the neutron excess can be evaluated 
in terms of the consequent contribution to 
$\la x_A\,u^-_A - x_A\,d^-_A\ra$, using Eq.~\eqref{eq:correction}.
For nuclei with $N > Z$ the $u$-quarks feel less vector repulsion than 
the $d$-quarks, and in Ref.~\cite{Cloet:2009qs} it was shown that a model
independent consequence of this is that there is a small shift in quark momentum 
from the $u$- to the $d$-quarks.
Therefore, the momentum fraction $\la x_A\,u^-_A - x_A\,d^-_A\ra$ in 
Eq.~\eqref{eq:correction} will be negative~\cite{Cloet:2009qs}, even after standard 
isoscalarity corrections are applied. Correcting for the isovector-vector field 
therefore has the \textit{model independent effect of reducing the NuTeV 
result for $\sin^2\theta_W$}.

To estimate the effect on the NuTeV experiment, Clo\"et \textit{et al.} 
used a nuclear matter approximation, chose the $Z/N$ ratio to 
correspond to the NuTeV experimental neutron excess and calculated
the quark distributions at an effective density appropriate
for Fe, namely 0.89 times nuclear matter density~\cite{Moniz:1971mt}. 
Using Eq.~(\ref{eq:correction}), this gave an estimate of the 
isovector correction of $\D R_{\text{PW}}^{\raisebox{1.0pt}{${\scriptstyle\rho}$}^0} = -0.0025$. 
Finally, the NuTeV CSV functional~\cite{Zeller:2002du} was used to
obtain an accurate determination of this effect on the NuTeV result. 
This gave $\Delta R^{\raisebox{1.0pt}{${\scriptstyle\rho}$}^0} = -0.0019$, 
which accounts for between 1.0 and 1.5$\sigma$ of the NuTeV discrepancy
with the Standard Model. 

The sign of this effect is model independent and because it depends only on 
the difference in the neutron and proton densities in iron and the 
symmetry energy of nuclear matter, which are both well known, the magnitude is 
expected to be well constrained.
As a conservative estimate of the uncertainty we assign 
an error twice that of the difference between the PW correction  
obtained at nuclear matter density and at 0.89 times that, this gives
\begin{equation}
\Delta R^{\raisebox{1.0pt}{${\scriptstyle\rho}$}^0} = -0.0019 \pm 0.0006.
\label{eq:rho_result}
\end{equation}

Other studies of nuclear corrections to the PW ratio have largely been 
focused on Fermi motion~\cite{Kulagin:2003wz,Cloet:2009qs} and nuclear 
shadowing~\cite{Kulagin:2003wz,Miller:2002xh,Brodsky:2004qa} effects. Fermi 
motion corrections were found to be small~\cite{Kulagin:2003wz,Cloet:2009qs} 
and the NuTeV collaboration
argue that, given their $Q^2$-cuts, sizeable corrections from 
shadowing would be inconsistent with
data~\cite{McFarland:2002sk}. Therefore we have not included a 
correction from shadowing here.

%===============================================================================
%###############################################################################
%===============================================================================
\textit{Charge Symmetry Violation} ---
Before the NuTeV result, two independent studies of the effect of quark mass 
differences on proton and neutron PDFs, 
by Sather~\cite{Sather:1991je} and by Rodionov \textit{et al.}~\cite{Rodionov:1994cg}, 
reached very similar conclusions. These mass differences  violate charge symmetry, 
the invariance of  the QCD Hamiltonian under a rotation by 180 degrees about 
the 2-axis in isospin space. They lead to the CSV differences 
\begin{align}
\delta d^-(x) &=  d^-_p(x) - u^-_n(x), \\
\delta u^-(x) &=  u^-_p(x) - d^-_n(x),  
\end{align}
where the subscripts $p$ and $n$ label the proton and neutron, respectively.
The contribution of CSV in the nucleon can be found through 
Eq.~\eqref{eq:correction} and has the form:
\begin{equation}
\D R^{\text{CSV}}_{\text{PW}} = 
\frac{1}{2}\left( 1 - \frac{7}{3} \sin^2 \theta_W \right) 
\frac{\left\la x\,\delta u^- - x\,\delta d^- \right\ra}
{\left\la x\,u^-_p + x\,d^-_p \right\ra}.
\label{eq:CSV_correction}
\end{equation}

Londergan and Thomas~\cite{Londergan:2003ij} explained the similarity of the 
results obtained by Sather and Rodionov \textit{et al.} by demonstrating that the 
leading contribution to the moment $\left\la x\,\delta u^- - x\,\delta d^- \right\ra$
is largely model independent and simply involves the ratio of the up-down 
mass difference to the nucleon mass. 
The contribution arising from the quark mass differences to Eq.~\eqref{eq:CSV_correction}
was found to be $\D R^{\delta m}_{\text{PW}} \simeq -0.0020$ and the corresponding NuTeV
CSV functional result was $\D R^{\delta m} \simeq -0.0015$~\cite{Londergan:2003ij}. 
We assign an error of 20\% to this term, which is conservative in view 
of its demonstrated model independence. 

An additional CSV effect arises from QED splitting~\cite{Gluck:2005xh,Martin:2004dh}, 
associated with the $Q^2$ evolution of photon emission from the quarks.
Because $\left|e_u\right| > \left|e_d\right|$ the $u$-quarks lose momentum to the
photon field at a greater rate than the $d$-quarks. 
Therefore a model independent consequence of QED splitting is that it will 
reduce the NuTeV result for $\sin^2\theta_W$.
Gl{\"u}ck \textit{et al.}~\cite{Gluck:2005xh} calculated this effect on the NuTeV result and obtained 
$\Delta R^{\text{QED}}_{\text{PW}} = -0.002$, corresponding to $\Delta R^{\text{QED}} = -0.0011$ using the NuTeV CSV functional. A similar study was 
undertaken by the MRST group~\cite{Martin:2004dh} who explicitly included QED 
splitting effects in the PDF evolution and found 
$\Delta R^{\text{QED}}_{\text{PW}} = -0.0021$ at $Q^2 = 20\,$GeV$^2$. This 
correction has the same sign 
as the CSV term arising from quark mass differences and the two contributions 
are almost independent so we simply add them. The sum of the two terms 
explains roughly half of the NuTeV discrepancy with the Standard Model. 
Assigning a conservative 100\% error to the QED splitting result and combining 
the errors in quadrature gives a total CSV correction of
\begin{equation}
\Delta R^{\text{CSV}} = -0.0026 \pm 0.0011.
\label{eq:CSV_result}
\end{equation}

The only experimental information regarding CSV effects on the PDFs is 
obtained from an MRST study~\cite{Martin:2003sk}. In this case a global 
analysis was performed on a set of high energy data allowing for explicit 
CSV in the PDFs. From their global analysis~\cite{Martin:2003sk} the MRST 
group found 
$\Delta R^{\text{CSV}}_{\text{PW}} = -0.002$, with a 90\% confidence 
interval of $-0.007 < \Delta R^{\text{CSV}}_{\text{PW}} < 0.007$.
The MRST study implicitly include both sources of CSV considered here, 
quark-mass and QED effects. The 90\% confidence interval obtained by MRST 
allows a rather large range of valence quark CSV. 

We have not adopted the MRST value and error for 
partonic CSV, for the following reasons. First, the MRST results are 
based on the assumption of a specific functional form for the CSV parton 
distributions. The assumed function had an overall strength parameter that 
was varied to obtain the best fit to the global analysis. MRST also 
imposed relations between the valence quark CSV PDFs. For convenience in 
their global analysis they neglected the $Q^2$ dependence of the CSV 
distributions. Finally, the experiments in the global set of high energy 
data have different treatments of radiative corrections. It is not clear 
that these different radiative corrections have been treated consistently 
in an analysis of CSV effects. If the CSV effects are as large as are 
allowed within the MRST 90\% confidence limit, it should be possible to 
observe such effects \cite{Londergan:2009kj}. However it will be some time 
before experiments can further constrain this result.

%===============================================================================
%###############################################################################
%===============================================================================

%===============================================================================
\begin{table*}[btp]
%\addtolength{\tabcolsep}{20pt}
\addtolength{\extrarowheight}{5.0pt}
\begin{ruledtabular}
\begin{tabular}{lccccc}
\\[-2.0em]
& $\la x\,s^-\ra$  & $\Delta R^s$ & $\Delta R^{\text{total}}$ & $\sin^2\theta_W$ $\pm$ syst. \\[0.1em]
\hline
Mason \EA~\cite{Mason:2007zz}     
& $0.00196 \pm 0.00143$  & $-0.0018 \pm 0.0013$  & $-0.0063 \pm 0.0018$ & $0.2214 \pm 0.0020$  \\
NNPDF~\cite{Ball:2009mk}             
& $0.0005 \pm 0.0086$    & $-0.0005 \pm 0.0078$  & $-0.0050 \pm 0.0079$ & $0.2227 \pm large$   \\
Alekhin \EA~\cite{Alekhin:2009mb} 
& $0.0013 \pm 0.0009 \pm 0.0002$ & $-0.0012 \pm 0.0008 \pm 0.0002$ &
$-0.0057 \pm 0.0015$ & $0.2220 \pm 0.0017$ \\
MSTW~\cite{Martin:2009iq}            
& $0.0016_{-0.0009}^{+0.0011}$ & $-0.0014_{+0.0008}^{-0.0010}$ &
$-0.0059 \pm 0.0015$ & $0.2218\pm 0.0018$ \\
CTEQ~\cite{Lai:2007dq} 
& $0.0018_{-0.0004}^{+0.0016}$ & $-0.0016_{-0.0004}^{+0.0014}$ &
$-0.0061_{-0.0013}^{+0.0019}$ & $0.2216_{-0.0016}^{+0.0021}$ \\
This work (Eq.~\eqref{eq:strange_result})
& $0.0 \pm 0.0020$ & $0.0 \pm 0.0018$ & $-0.0045 \pm 0.0022$ & $0.2232
\pm 0.0024$
\end{tabular}
\end{ruledtabular}
\caption{A summary of the recent estimates of the strangeness asymmetry, 
$\langle x s^- \rangle$, the correction to the Paschos-Wolfenstein
ratio after applying the NuTeV functional, $\Delta R^s$, and the total
correction, $\Delta R^{\text{total}}$,  obtained by combining $\Delta R^{\rho^0}$, 
$\Delta R^{\text{CSV}}$ and $\Delta R^s$, with the errors added in quadrature. 
The final column
shows the value of $\sin^2 \theta_W$ deduced in each case by applying
the total correction to the published NuTeV result.
Note that we show only the systematic error, which is obtained by treating the  
error on $\Delta R^{\text{total}}$ as a systematic error and
combining it in quadrature with the NuTeV systematic error.}
\label{tab:strange}
\end{table*}
%===============================================================================

\textit{Strange Quark Asymmetry} ---
A difference in shape between $s(x)$ and $\bar{s}(x)$ in the nucleon was 
first proposed on the basis of chiral symmetry in Ref.~\cite{Signal:1987gz}.
However, the size of $s^-(x)$ is not constrained by symmetries and badly 
needs further input from experiment or lattice QCD~\cite{Deka:2008xr}.
The strange quark correction arises from the term $\left\la x\,s_A^- \right\ra$
on the right hand side of Eq.~\eqref{eq:correction}.
The best direct experimental information on $\left\la x\,s_A^- \right\ra$ 
comes from opposite sign dimuon production in reactions induced by neutrinos 
or antineutrinos. Such experiments have been carried out by the 
CCFR~\cite{Bazarko:1994tt} and NuTeV~\cite{Goncharov:2001qe} groups. A
precise extraction of $\left\la x\,s_A^- \right\ra$ is the NuTeV analysis 
by Mason \EA~\cite{Mason:2007zz}, which found
$\left\la x\,s_A^- \right\ra = 0.00196 \pm 0.00143$ at $Q^2=16\,$GeV$^2$,
where we have added the various errors in quadrature. 

Global PDF analyses have also provided estimates of $s^-(x)$. The recent 
examination by the NNPDF collaboration found 
$\left\la x\,s^- \right\ra = 0.0005\pm 0.0086$~\cite{Ball:2009mk}
at $Q^2=20\,$GeV$^2$, which has an error more than six times larger than 
that cited by NuTeV.
However, unlike the NuTeV dimuon experiment, this analysis is not 
directly sensitive to the $s$-quark distributions, evidenced by the fact that
their upper limit on $s^-(x)$ is an order of magnitude 
larger than any other sea quark distribution at $x \sim 0.5$.  
This large uncertainty is a consequence of their neural network approach, 
which was primarily aimed at accurately determining $V_{cd}$ and $V_{cs}$, 
not $s^-(x)$~\cite{Ball:2009mk}. Alekhin \EA~\cite{Alekhin:2009mb} obtained 
$\la x\,s^- \ra = 0.0013 \pm 0.0009(\text{exp}) \pm 0.0002(\text{QCD})$ 
when they imposed a constraint on the semileptonic branching ratio $B_{\mu}$ 
from production rates of charmed hadrons in other experiments. 
The MSTW collaboration find a momentum fraction very similar to that of NuTeV, 
namely $\la x\,s^- \ra = 0.0016^{+0.0011}_{-0.0009}$~\cite{Martin:2009iq} 
at $Q^2=10\,$GeV$^2$, while CTEQ report $\left\la x\,s^- \right\ra = 0.0018$~\cite{Lai:2007dq} 
at $Q^2=1.69\,$GeV$^2$, with a 90\% confidence interval of 
$-0.001 < \left\la x\,s^- \right\ra < 0.005$. These results are summarized in
the second column of Table~\ref{tab:strange}.

The quantity $s_A^-(x)$ must have at least one zero-crossing since its first
moment vanishes. For each of the above analyses the central best fit curve 
crosses zero at values of $x$ less than $0.03$ for $Q^2 > 2\,$GeV$^2$, with
the exception of the NNPDF result which has a zero-crossing at $x = 0.13$ for $Q^2 = 2\,$GeV$^2$.
For example the NuTeV result has the zero-crossing at $x=0.004$, which
is a very small $x$ value 
(it is smaller than the lowest $x$ point measured in the CCFR and NuTeV 
experiments), and moreover is extremely unlikely on theoretical 
grounds~\cite{Signal:1987gz,Melnitchouk:1999mv,Ding:2004ht}. 
In any quark model 
calculation the zero-crossing will occur near $x \simeq 0.15$ (a value similar
to that found by NNPDF). In this case the NuTeV strange quark momentum fraction becomes 
$\left\la x\,s_A^- \right\ra = 0.00007$~\cite{Mason:2007zz}, with a 
moderate increase in the $\chi^2$ compared to the best value of Mason \EA. 
Since relatively little is known about the $s$-quark 
distributions we ignore nuclear effects and therefore assume 
$\left\la x\,s_A^- \right\ra \equiv \left\la x\,s^- \right\ra$ throughout
this discussion.

Clearly the correction to the PW ratio from the strange quark asymmetry has a 
significant uncertainty. On the theoretical grounds 
explained earlier, we prefer the 
NuTeV analysis based on a zero-crossing at $x \approx 0.15$, which means that 
$\left\la x\,s^- \right\ra$ is essentially zero. For the uncertainty we choose 
the difference between this and the NuTeV determination 
noted above with the zero-crossing at $x = 0.004$, this gives 
$\la x\,s^- \ra =  0.0 \pm 0.0020$ at 16 GeV$^2$.
This is a conservative choice for the error since it is substantially larger 
than the original uncertainty quoted by NuTeV and covers all of the central 
values of the analyses mentioned above. 
%Including the effect of the NuTeV 
%functional leads to the final input to our evaluation of the revised NuTeV 
%value of $\sin^2 \theta_W$, namely
Including the effect of the NuTeV functional leads to our preferred value 
for the strange quark correction to the NuTeV $\sin^2 \theta_W$ result, namely
\begin{align}
\Delta R^s = 0.0 \pm 0.0018.
\label{eq:strange_result}
\end{align}
The $s$-quark corrections to the NuTeV result obtained from the other analyses
discussion here are summarized in column three of Table~\ref{tab:strange}.

%===============================================================================
%###############################################################################
%===============================================================================
\textit{Conclusion} ---
The errors associated with the three corrections given in 
Eqs.~\eqref{eq:rho_result}, \eqref{eq:CSV_result} and \eqref{eq:strange_result} 
are systematic and to a very good approximation independent errors. 
We therefore combine them in quadrature with 
the original systematic error quoted 
by NuTeV. The statistical error is, of course, 
unchanged from the NuTeV analysis. 

Because of the uncertainty over the strangeness asymmetry, in the last column
of Table~\ref{tab:strange} we show the effect on $\sin^2 \theta_W$ for each of the recent 
analyses~\cite{Mason:2007zz,Ball:2009mk,Alekhin:2009mb,Martin:2009iq,Lai:2007dq}
as well as our own preferred value given in
Eq.~(\ref{eq:strange_result}).
Every one of the six results lies within one standard deviation of 
the Standard Model value for $\sin^2\theta_W$.
As a best estimate of the corrected value we take
the average of these six values. For the systematic error we note that 
(apart from NNPDF which is unrealistically large) they are all very 
similar. Because of the correlations between them, the final quoted
systematic error is a simple average of all the entries in the 
last column of Table~\ref{tab:strange} except NNPDF. This yields the revised value 
for $\sin^2 \theta_W$, including all of the corrections discussed 
here, namely: 
\begin{align}
\sin^2 \theta_W &= 0.2221 \pm 0.0013(\text{stat}) \pm 0.0020(\text{syst}), 
\label{eq:final_result}
\end{align}
which is in excellent agreement with the Standard Model expectation
of $\sin^2 \theta_W = 0.2227 \pm 0.0004$~\cite{Abbaneo:2001ix,Zeller:2001hh}.
Correction terms of higher order than Eq.~\eqref{eq:correction} and also
$\mathcal{O}(\alpha_s)$ corrections were also investigated 
and found to be negligible.

\begin{figure}[tbp]
\centering\includegraphics[width=\columnwidth,clip=true,angle=0]{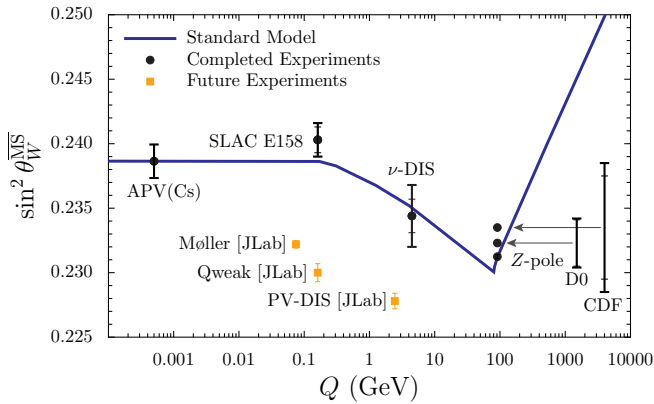}
\caption{The curve represents the running of $\sin^2\theta_W$ in the $\overline{\text{MS}}$
renormalization scheme~\cite{Erler:2004in}. 
The $Z$-pole point represents the combined results of six LEP and
SLC experiments~\cite{:2005ema}. 
The CDF~\cite{Acosta:2004wq} and D0~\cite{:2008xq} collaboration results
(at the $Z$-pole) and the SLAC E158~\cite{Anthony:2005pm} result,
are labelled accordingly.
The atomic parity violating (APV) result~\cite{Porsev:2009pr}
has been shifted from $Q^2\to 0$ for clarity.
The inner error bars represent the statistical uncertainty 
and the outer error bars the total uncertainty. 
At the $Z$-pole, conversion to the 
$\overline{\text{MS}}$ scheme was achieved via 
$\sin^2\theta_W^{\text{eff}} = 0.00029 + \sin^2\theta_W^{\overline{\text{MS}}}$~\cite{:2005ema}.
For the results away from the $Z$-pole, the discrepancy with the Standard Model 
curve reflects the disagreement with the Standard Model in the renormalization
scheme used in the experimental analysis.
}
\label{fig:sin2tw}
\vspace{-1.0em}
\end{figure}

This updated value for the NuTeV determination of $\sin^2 \theta_W$ is 
also shown in  Fig.~\ref{fig:sin2tw}, now in the 
$\overline{\text{MS}}$-scheme and labelled as $\nu$-DIS, 
along with the results of a number of other completed experiments and 
the anticipated errors of several future experiments, which are shown at the
appropriate momentum scale $Q$. 

In this paper we have summarized various estimates of the size of both 
partonic CSV effects and a possible strange quark momentum asymmetry. For 
valence quark CSV we have relied on well founded 
theoretical arguments to constrain 
the magnitude of CSV effects arising from quark mass differences. We 
have also used theoretical guidance on the zero crossing in $s^-(x)$, 
as well as the most recent analyses of the  
experimental data to constrain 
the strange quark momentum asymmetry. 
When re-evaluated, the NuTeV point is within one standard 
deviation of the Standard Model prediction for all analyses of 
this asymmetry. 
As the experimental information on the strange quark asymmetry or 
charge symmetry violation improves it is a simple matter to update the 
current analysis.

%===============================================================================
%===============================================================================

As a final point, we return to the fact that the NuTeV experiment actually 
measured $R^\nu$ and $R^{\bar{\nu}}$, not $R_{\rm PW}$. We might ask  
how the corrections that we have applied affect the individual values 
for these two ratios. For the quantity $R^\nu$ NuTeV measured 
$0.3916 \pm 0.0013$, compared with 0.3950 in the Standard Model, while for 
$R^{\bar{\nu}}$ they obtained $0.4050 \pm 0.0027$, compared with 0.4066. 
The corrections to the Standard Model ratios arising from the 
isovector EMC effect and CSV are both included through the non-zero value 
of $\la x_A\, u_A^- - x_A\, d_A^- \ra$:
%
%\begin{align}
%\delta R^{\nu} &= \frac{2 (g_{Lu}^2 - g_{Ru}^2/3) \la x u_A^- - xd_A^- \ra} 
%{\la x u_A +x d_A + (x \bar{u}_A + x \bar{d}_A)/3 + 2 x s_A \ra},  \\
%\delta R^{\bar{\nu}} &=  \frac{-2(g_{Rd}^2 - g_{Ld}^2/3) \la x u_A^- - x d_A^- \ra}{\la (x u_A + x d_A)/3 + x \bar{u}_A + x \bar{d}_A + 2 x \bar{s}_A\ra}.
%\label{eq:Rcorrns}
%\end{align}
%
\begin{align}
\delta R^{\nu} &= \frac{2\lf(3\,g_{Lu}^2 + g_{Ru}^2\rg) \lf\la x_A\,u_A^- - x_A\,d_A^- \rg\ra} 
{\lf\la 3\,x_A\,u_A + 3\,x_A\,d_A + x_A\,\bar{u}_A + x_A\,\bar{d}_A + 6\,x_A\, s_A \rg\ra},  \\
\delta R^{\bar{\nu}} &=  \frac{-2\lf(g_{Ld}^2 + 3\,g_{Rd}^2\rg) \lf\la x_A\,u_A^- - x_A\,d_A^- \rg\ra}
{\lf\la x_A\,u_A + x_A\,d_A + 3\,x_A\,\bar{u}_A + 3\,x_A\,\bar{d}_A + 6\,x_A\,\bar{s}_A\rg\ra},
\label{eq:Rcorrns}
\end{align}
where
\begin{align}
g_{Lu} &= \phantom{-}\frac{1}{2} - 
\frac{2}{3}\sin^2\theta_W & g_{Ru} &= -\frac{2}{3}\sin^2\theta_W, \\
g_{Ld} &= -\frac{1}{2} + 
\frac{1}{3}\sin^2\theta_W           & g_{Rd} &= \phantom{-}\frac{1}{3}\sin^2\theta_W.
\end{align}
It is clear from our earlier discussion 
that $\la x_A\, u_A^- - x_A\, d_A^- \ra $ 
is negative and allowing for the NuTeV functional 
we find $\delta R^\nu = -0.0017 \pm
0.0008$ and $\delta R^{\bar{\nu}} = +0.0016 \pm 0.0008$. 
(Note that the errors quoted 
also include our estimated error on $\la x s^- \ra$). 
Subtracting 
$\delta R^\nu$ from the NuTeV result yields a 
value $0.3933 \pm 0.0015$, which is in 
good agreement with the Standard Model value, namely 0.3950. 
The corresponding 
$\bar{\nu}$ correction yields $R^{\bar{\nu}} = 0.4034 \pm 0.0028$, 
which is just over 
one standard deviation from the Standard Model value, namely 0.4066 - 
again, in quite good agreement. After including our corrections from 
nuclear effects, partonic CSV and strange quarks, the total $\chi^2$ for 
$R^\nu$ and $R^{\bar{\nu}}$ compared with the Standard Model values moves from 
7.19 to 2.58.
This represents a very significant improvement.

In conclusion, it should be clear that there is no longer any significant  
discrepancy between the predictions of the Standard Model evolution and 
the existing data. However, we look forward to the much higher accuracy 
in the weak mixing angle which is anticipated in future 
experiments \cite{Opper:2008zz,Souder:2008zz,Kumar:2009zz}. With regard 
to NuTeV itself, the greatest single improvement in the accuracy with 
which one could extract $\sin^2 \theta_W$ would come from a more precise 
determination of  $\left\la x\,s^- \right>$. A decrease in the
error associated with $R^{\bar{\nu}}$ would also set tight constraints 
on QCD corrections to the NuTeV result.

\begin{acknowledgments}
We would like to thank Jens Erler, Larry Nodulman and Heidi Schellman for 
helpful correspondence. We also thank W.~van~Oers and R.~Young for useful discussions.
This work was supported by the Australian Research Council through an Australian 
Laureate Fellowship (AWT), by the U.S. Department of Energy under Grant No. DEFG03-97ER4014
and by Contract No. DE-AC05-06OR23177, under which Jefferson Science Associates,
LLC operates Jefferson Laboratory, by the Grant in Aid for Scientific
Research of the Japanese Ministry of Education, Culture, Sports, Science and
Technology, Project No. C-19540306 and by U.S. National Science Foundation 
grant NSF PHY-0854805.
\end{acknowledgments}

%===============================================================================
%###############################################################################
%===============================================================================
%\vspace{-1em}


\begin{thebibliography}{50}
\bibitem{paschos}
  The cross-sections in Eq.~\eqref{eq:PW} have been integrated over the Bjorken scaling
  variable and energy transfer.
\end{thebibliography}
\end{document}